\documentclass[prb,aps,superscriptaddress,amsmath,amssymb,twocolumn,reprint]{revtex4-2}
\usepackage{CJKutf8}
\usepackage{graphicx}
\usepackage{dcolumn}
\usepackage{bm,braket}

\usepackage{mathptmx}
\usepackage{graphicx}
\usepackage{subfigure}
\usepackage{amsmath}
\usepackage{amsfonts}
\usepackage{amssymb}
\usepackage{mathrsfs}
\usepackage{subfigure}
\usepackage{xcolor}
\usepackage[colorlinks=true,linkcolor=blue,anchorcolor=red,citecolor=blue, urlcolor=blue,breaklinks=true]{hyperref}
\usepackage{ulem}
\usepackage{pifont}

\makeatletter
\renewcommand*\env@matrix[1][\arraystretch]{%
  \edef\arraystretch{#1}%
  \hskip -\arraycolsep
  \let\@ifnextchar\new@ifnextchar
  \array{*\c@MaxMatrixCols c}}
\makeatother

\def\be{\begin{equation}}
\def\ee{\end{equation}}
\def\ba{\begin{eqnarray}}
\def\ea{\end{eqnarray}}

\newlength{\seplinewidth}
\newlength{\seplinesep}
\colorlet{sepline}{orange}

\begin{document}
\begin{CJK*}{UTF8}{gbsn}

\title{Escaping Local Minima with Quantum Coherent Cooling}

\author{Jia-Jin Feng(冯嘉进)}

\affiliation{International Center for Quantum Materials, School of Physics, Peking University, Beijing 100871, China}
\affiliation{Ming Hsieh Department of Electrical and Computer Engineering, University of Southern California, Los Angeles, California 90089, USA}

\author{Biao Wu(吴飙)}
\email{wubiao@pku.edu.cn}
\affiliation{International Center for Quantum Materials, School of Physics, Peking University, Beijing 100871, China}
\affiliation{Wilczek Quantum Center, School of Physics and Astronomy, Shanghai Jiao Tong University, Shanghai 200240, China}
\affiliation{Collaborative Innovation Center of Quantum Matter, Beijing 100871, China}

\date{\today}
\begin{abstract}
Quantum cooling has demonstrated its potential in quantum computing, which can reduce the number of control channels needed for external signals. Recent progress also supports the possibility of maintaining quantum coherence in large-scale systems. The limitations of classical algorithms trapped in local minima of cost functions could be overcome using this scheme. According to this, we propose a hybrid quantum-classical algorithm for finding the global minima. Our approach utilizes quantum coherent cooling to facilitate coordinative tunneling through energy barriers if the classical algorithm gets stuck. The encoded Hamiltonian system represents the cost function, and a quantum coherent bath in the ground state serves as a heat sink to absorb energy from the system. Our proposed scheme can be implemented in the circuit quantum electrodynamics (cQED) system using a quantum cavity. The provided numerical evidence demonstrates the quantum advantage in solving spin glass problems.
\end{abstract}
\maketitle
\end{CJK*}

Optimization problems are prevalent in many areas, where the global minima of cost functions represent the optimal solutions. Classical algorithms are usually able to quickly identify local minima but get trapped due to the complexity of the cost function. As a result, they are inefficient at finding the global minima. This bottleneck is widely recognized in various fields, including computational physics and machine learning.

\begin{figure}[b]
\begin{center}
\includegraphics[clip = true, width =\columnwidth]{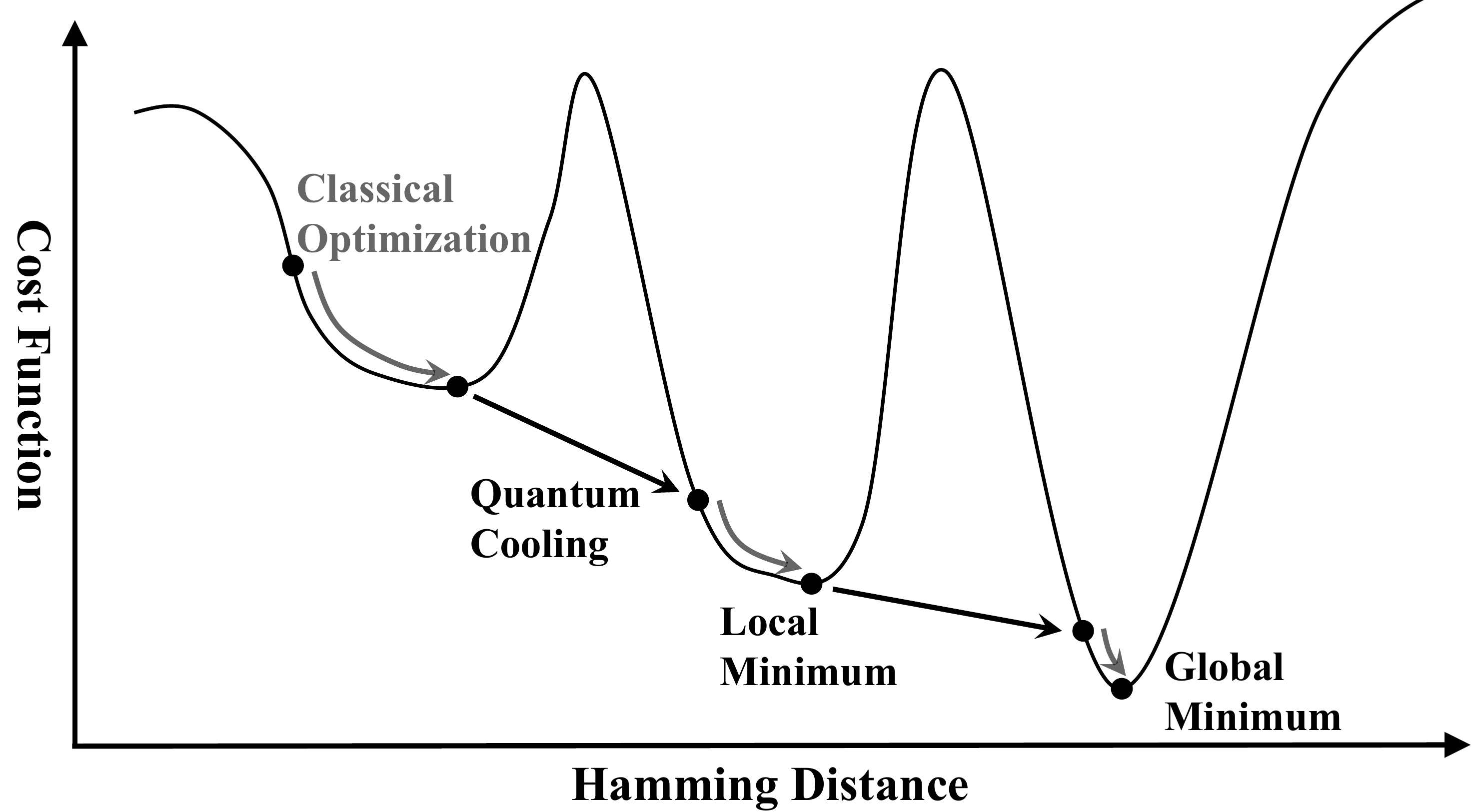}
\caption{\label{fig:potential} The proposed heterogeneous cooling process. }
\end{center}
\end{figure}

We proposed a hybrid quantum-classical algorithm to address this bottleneck. The entire algorithm functions as a heterogeneous cooling process to minimize the value of the cost function that encodes the given problem. The classical optimization is employed to find local minima, while the quantum cooling helps to escape the local minima by tunneling through energy barriers \cite{Espinosa2019}. By alternating between classical optimization and quantum cooling, one can ultimately reach the global minimum. (See Fig.\ref{fig:potential} for a visualization of this process.)

In the classical part of the algorithm, well-established classical techniques, such as Monte Carlo or gradient descent, are used to identify a local minimum of the cost function with minor computational resources. In the quantum part of the algorithm, the quantum icebox algorithm (QIA) performs the quantum cooling, which has been shown to achieve quantum advantage in the unsorted search problem \cite{Feng2022}. The framework of QIA is as follows. The cost function is encoded in a Hamiltonian system \cite{Farhi2001,Ebadi2022}, which is initialized in the state corresponding to the local minimum. A quantum bath has a trivial and easy-to-prepare ground states that is initialized within it. The system and the bath are coupled and maintain coherence, which facilitates coordinative quantum tunneling over high-energy barriers. The system will evolve to escape the local minimum and transition to a lower value. These two parts are complementary in order to prevent the algorithm from getting stuck.

Other hybrid quantum-classical algorithms
\cite{Farhi2014,Peruzzo2014,McClean2016,Colless2018,McClean2018,McArdle2019,Zhou2020,Cerezo2021,Endo2021,Xin2021,Wang2022} also combine classical and quantum methods as
a compromise in the noisy intermediate-scale quantum (NISQ) era \cite{Preskill2018,Bharti2022},
such as the variational quantum algorithm (VQA).  In VQA, classical algorithms optimize the quantum gate parameters to reduce the number of quantum operations during the coherence time \cite{Cao2019}. However, this approach faces challenges such as the barren plateau \cite{McClean2018,Elies2023} and the need to avoid local minima. In our scheme, the classical algorithm only needs to optimize the initial state and find local minima, which is less problematic. Furthermore, control channels are not required during QIA, which helps to minimize noise from external sources.

We specifically target a category of problems where the cost function involves $n$ Boolean
variables \cite{Bollig1996,Kolmogorov2004}, such as the 3-SAT problem and independent sets \cite{TOVEY1984,SAKAI2003}. In physical terms, these cost functions represent Hamiltonians of (pseudo-)spins \cite{Hu2021,Yu2021}. The configuration space for these problems consists of binary strings of zeros and ones (or $\pm1$ for spins), with each string representing a state $\psi_{\rm s}$. Local minima for this class of problems are defined as states where neighboring states within one Hamming distance have higher cost function values (or energies). The Hamming distance is calculated by counting the number of different bits between two binary strings \cite{Trugenberger2001}. To solve these problems, our hybrid quantum-classical algorithm operates as follows (see also Fig. \ref{fig:potential}) :
\begin{enumerate}
\item Use arbitrary configuration $\psi_{\rm s0}$ as the initial condition to find out one of the local minima $\psi_{\rm s1}$ with classical algorithms.

\item Prepare the quantum state $\psi_{\rm s1}$ in the computational basis, denoted as $|\psi_{\rm s1}\rangle$, and execute the QIA. Following quantum evolution, we obtain an entangled state $|\Psi\rangle$. We then proceed to measure the problem system in the computational basis, resulting in the random selection of a configuration denoted as $\psi_{\rm s2}$.

\item Use the configuration $\psi_{\rm s2}$ as the initial condition to identify a different local minimum, represented as $\psi_{\rm s3}$, using classical algorithms.

\item If $\langle \psi_{\rm s3}| H_{\rm s} |\psi_{\rm s3}\rangle\leqslant\langle \psi_{\rm s1}| H_{\rm s} |\psi_{\rm s1}\rangle$, then we update the initial configuration from $\psi_{\rm s1}$ to $\psi_{\rm s3}$. In any case, we repeat the process starting from step 2.

\item Stop the iteration if $\langle\psi_{\rm s1}| H_{\rm s} |\psi_{\rm s1}\rangle$ is sufficiently small and no further update to $\psi_{\rm s1}$ is available.
\end{enumerate}

This hybrid quantum-classical algorithm is robust to noise and errors. The long Hamming distance tunneling is divided into several shorter Hamming distance tunnelings among local minima, making it suitable for qubits with limited coherence time. Furthermore, it can identify errors in step 4 and correct them through multiple rounds of optimization. These properties make it feasible to apply them practically and implement them experimentally.

\begin{figure}[tb]
\begin{center}
\includegraphics[clip = true, width =\columnwidth]{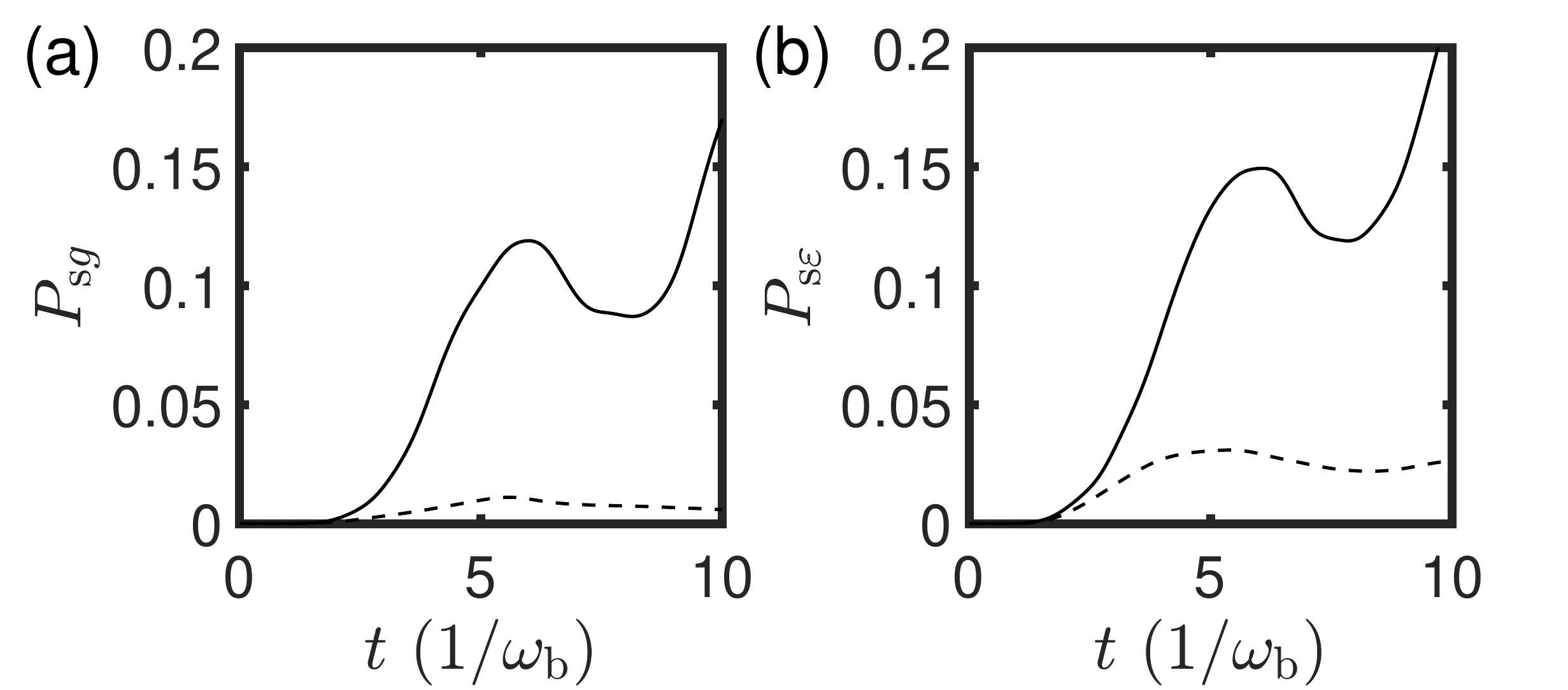}
\caption{\label{fig:Pg} Time evolution of the system cooled by the quantum bath (solid line) or the classical bath (dashed line). The number of qubits is $n_{\rm s}=11$ $(2^{-n_{\rm s}}=4.9\times10^{-4})$. The interaction strength is $\lambda=0.2\hbar\omega_{\rm b}$. The initial state of the problem system is the high energy local minimum. (a) The probability of the ground state of the problem system (global minimum). (b) The total probability of states of the problem system with lower on-site energy than the initial state.}
\end{center}
\end{figure}

With the assistance of classical algorithms, it can efficiently reduce local energy. This hybrid algorithm achieves quantum acceleration in step 2 by utilizing the quantum bath for efficient global cooling. We can also implement continuous quantum error correction \cite{Ahn2003,Livingston2022} or insert discrete quantum error correction \cite{Fowler2012,Bluvstein2023} in step 2 to suppress errors.
We will demonstrate that using the quantum bath leads to higher cooling efficiency.

Since classical algorithms are well-established, let us now focus on the quantum component. We will compare the cooling efficiency of the quantum coherence bath with that of the classical bath to demonstrate the quantum advantage, particularly in the context of spin glasses. Spin glasses are typically challenging to cool down because they often have numerous local minima. Our numerical calculations demonstrate that the quantum coherence bath has a faster cooling speed and a higher success rate in comparison to the classical bath (see Fig. \ref{fig:Pg}).


To achieve cooling, ancilla interacting qubits are commonly used as the quantum bath \cite{Feng2022,Matthies2022,Briones2016,Zaiser2021}. However, this approach often requires at least twice the number of qubits \cite{Feng2022,Matthies2022}, and maintaining entanglement among a large number of qubits is difficult in the NISQ era \cite{Preskill2018,Bharti2022}. As an alternative, we use a coherent cavity as the quantum bath, which is more feasible than using hundreds of qubits. The cavity is typically used in experiments to propagate interactions \cite{Majer2007,Blais2007} or measure states \cite{Blais2021}, and can take the form
of a laser cavity \cite{Gershon2015,Aspelmeyer2014}, inductor-capacitor ($LC$) oscillator \cite{Blais2021},
or nanomechanical resonator \cite{Cleland2004}.

\begin{figure}[tb]
\begin{center}
\includegraphics[clip = true, width =\columnwidth]{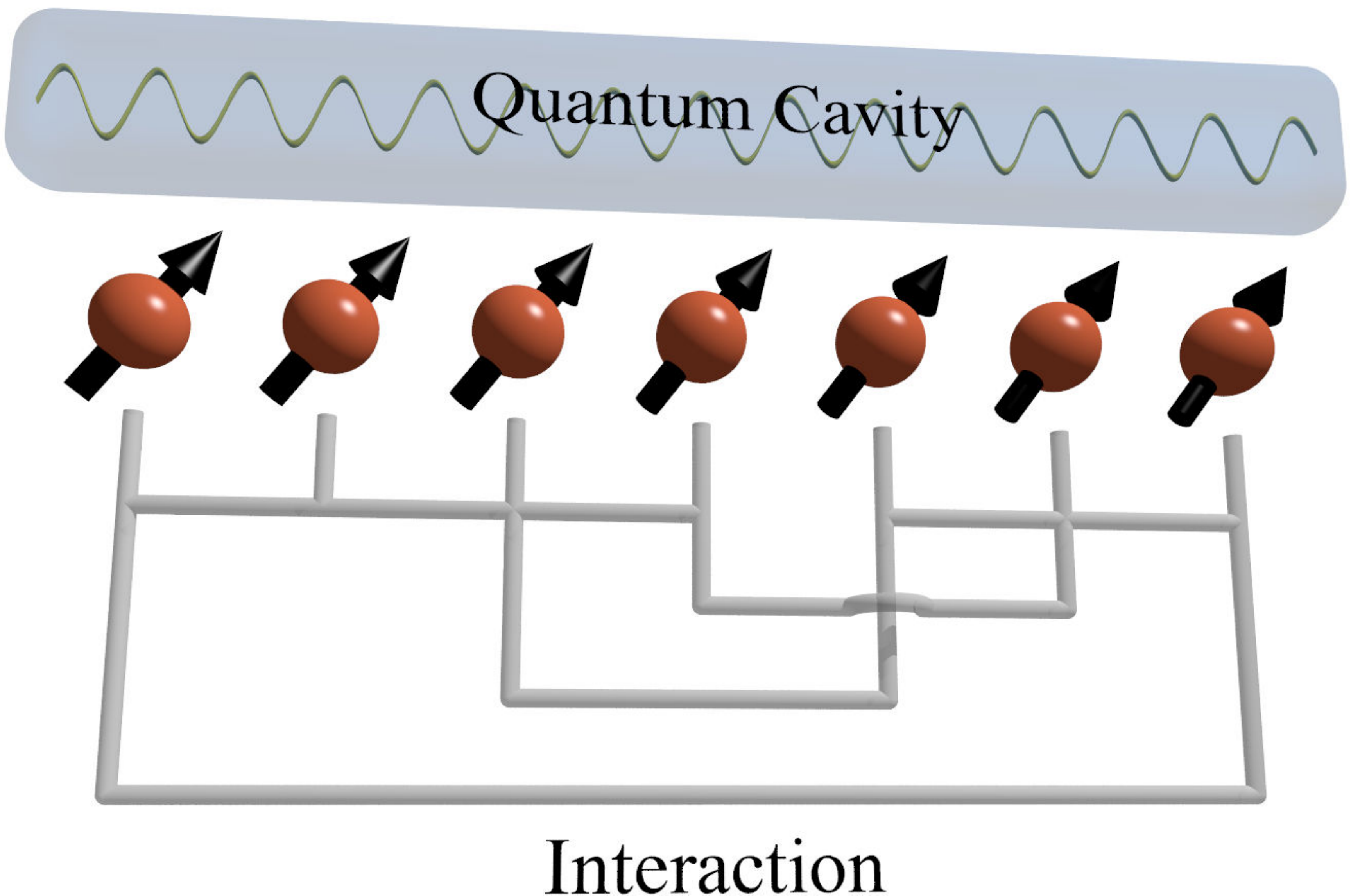}
\caption{\label{fig:setup} (color online) The setup of our system. The coupled qubits encode the cost function and the quantum cavity serves as the bath.}
\end{center}
\end{figure}

Our discussion will focus on the circuit quantum electrodynamics (cQED) systems \cite{Blais2021},
although our algorithm can also be applied to other systems. The basic setup is depicted in Fig. \ref{fig:setup}, where the cavity in the superconductor circuit can be modeled as an $LC$ oscillator. The angular frequency of the oscillator is given by $\omega_{\rm b}=1/\sqrt{LC}$, where $L$ is the inductance and $C$ is the capacitance. The quantum cavity acts as the bath, and its Hamiltonian is
\begin{eqnarray}
H_{\rm b} &=& \hbar\omega_{\rm b} \left(\hat{b}^\dagger \hat{b}+\frac{1}{2} \right),
\label{eq:Hb}
\end{eqnarray}
where $\hat{b}^\dagger$ is the creator of the harmonic mode of the bath.

The qubits in Josephson junctions store the problem data, and their interactions are established through circuit connections. The Hamiltonian encoding the problem of $n_s$ qubits is given by
\cite{Collodo2020,Lucas2014,Zhao2020,MingGong2021,King2022}
\begin{eqnarray}
H_{\rm s} &=& \sum_{m=0}^{n_{\rm s}-1}J_{m}^{(1)}\hat{\sigma}_m^z+\sum_{\langle m,m'\rangle}J_{m,m'}^{(2)}\hat{\sigma}_m^z\hat{\sigma}_{m'}^z ,
\label{eq:Hs}
\end{eqnarray}
which describes a spin glass that is difficult to cool down. The on-site energy is denoted by $J_m^{(1)}$, while $J_{m,m'}^{(2)}$ represents the interaction between two qubits. This interaction can be spatially non-local, making the optimization problem more challenging. To facilitate quantum transitions, $J_m^{(1)}$ and $J_{m,m'}^{(2)}$ are multiples of $\omega_{\rm b}/2$. The occurrence of integer multiple gaps is common when encoding discrete mathematical problems, including non-deterministic polynomial (NP) problems \cite{Farhi2001}.  It also helps to suppress overflow errors caused by unencoded energy levels.

The interaction between the qubits and the cavity in the lumped-element circuit without spatial dependence is given by \cite{Gely2017,Malekakhlagh2017,Blais2021,Miyanaga2021,Altoe2022}
\begin{eqnarray}
H_{\lambda} &=& \lambda \left(\hat{b}^\dagger- \hat{b}\right)\sum_{m=0}^{n_{\rm s}-1}\left(\hat{\sigma}_m^+-\hat{\sigma}_m^-\right),
\label{eq:Hi}
\end{eqnarray}
where $\lambda$ is the interaction strength proportional to the capacitance. The interconnected capacitor introduces additional on-site capacitance, which can be absorbed in the parameters of $H_{\rm s}$ and $H_{\rm b}$. Our scheme involves that  $\langle H_\lambda \rangle$ is comparable to $\langle H_{\rm s} \rangle$, resulting in acceleration and novel phenomena beyond perturbation approach, including rotating wave approximation (RWA) \cite{Kuzmin2019,Talkner2020,Liu2021}.

The cooling process is governed by the fixed total Hamiltonian $H=H_{\rm s}+H_{\rm b}+H_{\lambda}$ with little additional control. The circuit diagram is available in Appendix \ref{app:circuit}. The total parity conservation $e^{i\pi\left(b^\dagger b+0.5\sum_{m=0}^{n_{\rm s}-1}\hat{\sigma}_m^z \right)}$ can slightly simplify the numerical simulation.

The initial state of the bath is its ground state \cite{Mark2016,Schuster2007}, and the initial state
of the problem system is set to be one of the high energy local minima of $H_{\rm s}$ in the $\sigma_z$-basis.
Subsequently, the interaction $H_\lambda$ is turned on, and the state evolves according
to the Schr\"odinger equation. The energy of the system is absorbed by the bath through quantum tunneling,
leading to the transfer of the system's state to other local minima with lower energy.

For comparison, we treat the cavity as a classical system by defining the generalized momentum
 $\hat{Q}_{\rm b}=i\sqrt{\hbar\omega_{\rm b} C/2}\left(\hat{b}^\dagger-\hat{b}\right)$
 and the generalized position
 $\hat{\Phi}_{\rm b}=\sqrt{\hbar/2\omega_{\rm b} C}\left(\hat{b}^\dagger+\hat{b}\right)$ \cite{Blais2021}.
  If the decoherence is strong, we can treat them as classical variables $Q$ and $\Phi$
  \cite{Caldeira1983,Anatoli2010,Curtright2012,Wang2021}.
 The Hamiltonians for the bath in Eq. (\ref{eq:Hb}) and the interaction in Eq. (\ref{eq:Hi}) can
 be rewritten, respectively, as
 \begin{eqnarray}
H_{\rm b} &=& \frac{Q_{\rm b}^2}{2C_{\rm b}}+\frac{\Phi_{\rm b}^2}{2L_{\rm b}}\,,
\end{eqnarray}
and
\begin{eqnarray}
H_{\lambda} &=& \sqrt{\frac{2}{\hbar\omega_{\rm b} C_{\rm b}}}\lambda Q_{\rm b}\sum_{m=0}^{n_{\rm s}-1}\hat{\sigma}_m^y .
\end{eqnarray}
With the canonical equations $\partial Q_{\rm b}/\partial t=\partial \langle H \rangle/\partial \Phi_{\rm b}$ and $\partial \Phi_{\rm b}/\partial t=\partial \langle H \rangle/\partial Q_{\rm b}$ \cite{Zhangqi},
we derive the total dynamical equations as
\begin{eqnarray}
    i\hbar\frac{\partial\left| \psi_{\rm s} \right\rangle}{\partial t}&=&\left(H_{\rm s}+\sqrt{\frac{2}{\hbar\omega_{\rm b} C_{\rm b}}}\lambda Q_{\rm b}\sum_{m=0}^{n_{\rm s}-1}\hat{\sigma}_m^y \right) \left| \psi_{\rm s} \right\rangle ,  \\
    \frac{\partial Q_{\rm b}}{\partial t}&=&-\frac{\Phi_{\rm b}}{L_{\rm b}} ,\\
     \frac{\partial \Phi_{\rm b}}{\partial t}&=&\frac{Q_{\rm b}}{C_{\rm b}} +\sqrt{\frac{2}{\hbar\omega_{\rm b} C_{\rm b}}}\lambda \sum_{m=0}^{n_{\rm s}-1}\left\langle\hat{\sigma}_m^y\right\rangle .
\end{eqnarray}
Notice that $Q_{\rm b}=\Phi_{\rm b}=0$ is the fixed point if all the qubits are aligned in the $\sigma_z$ direction.
The classical bath can be treated as a classical probability ensemble. For a fair comparison,
the initial state of $Q_{\rm b}$ and $\Phi_{\rm b}$
is set to have the same probability distribution as the quantum ground state, which is $\rho_{\rm c}=e^{-\left(\phi_{\rm b}^2+q_{\rm b}^2\right)}/\pi$ with $q_{\rm b}=Q_{\rm b}/\sqrt{2\hbar\omega_{\rm b} C_{\rm b}}$
and $\phi_{\rm b}=\Phi_{\rm b}\sqrt{\omega_{\rm b} C_{\rm b}/(2\hbar)}$ \cite{Anatoli2010,Curtright2012}.
The expectation value is considered as the average of all the independent evolution paths,
i.e., $A=\iint a*\rho_{\rm c}{\rm d }q_{\rm b}{\rm d }\phi_{\rm b}$, where $a$ is the observable for one initial condition and $A$ is the expectation value.

No entanglement exists between the classical bath and the system, and the initial state
of the bath $\rho_{\rm c}$ can be regarded as a mixed state. Throughout the evolution,
the formation of entanglement is prohibited, and the total system remains in the product state $\left| \psi_{\rm s} \right\rangle\otimes\left| Q,\Phi \right\rangle$. Moreover, some high energy paths of the classical bath are also not allowed, even though they could help the problem system surmount barriers. The numerical results depicted in Fig. \ref{fig:Pg} demonstrate that the classical bath performs worse than its quantum counterpart.

We simulate a problem system with two local minima using Eq. (\ref{eq:Hs}). The initial state
of the problem system is at the higher local minimum, while the bath is in its ground state.
The Hamming distance between the local minima is $\lceil n_{\rm s}/2\rceil$, which increases the problem complexity. Cooling with the quantum bath does not always outperform the classical bath,
but quantum resonance can strongly accelerate cooling when $\lambda$ is
tuned suitably \cite{Childs2004}. Fig. \ref{fig:Pg} shows the probability $P_g$ of the ground state
changing over time. Cooling with the classical bath exhibits a modest increase over time,
while the curve for the quantum bath shows a sharp increase due to quantum coherent enhancement. It indicates that QIA has an advantage in escaping local minima.

This quantum acceleration is related to the entanglement and correlation numerically calculated in Appendix \ref{app:SC}.
In the continuous variable system, we also find a similar quantum acceleration (see Appendix \ref{app:Continous}).

\begin{figure}[tb]
\begin{center}
\includegraphics[clip = true, width =\columnwidth]{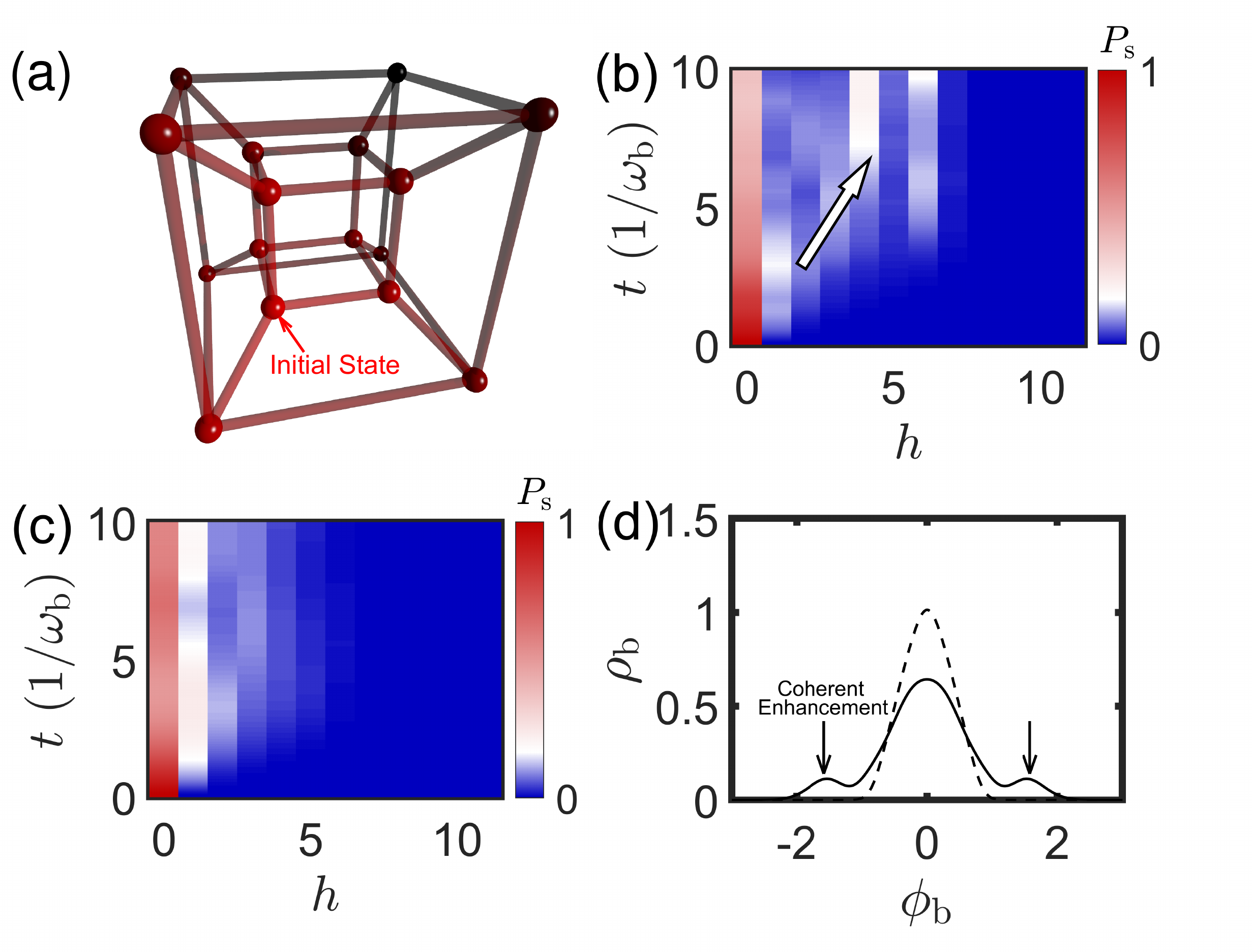}
\caption{\label{fig:diffusion} (color online) (a) The hypercube where each vertex represents an eigenstate of the spin glass.
The system is initially at the light red point, which is one of the local minima. The time evolution of
the probability distribution as the problem system is cooled by (b) the quantum bath and (c) the classical bath. $h$ is the Hamming distance from the initial state;
the value at each $h$ is the addition of probabilities of different states with the same $h$.
(d) The probability distribution of the quantum bath (solid line) and the classical bath (dashed line)
at $t=6/\omega_{\rm b}$.
The parameters used in the numerical computation are the same as those in Fig. \ref{fig:Pg}.}
\end{center}
\end{figure}

The eigenstates of the spin glass problem system $H_{\rm s}$ are represented by a hypercube
whose vertices correspond to the eigenstates in the $\sigma_z$ basis as shown in Fig. \ref{fig:diffusion}(a).
When the interaction is turned on, the wave function diffuses in the hypercube. As depicted in Fig. \ref{fig:diffusion}(b), the diffusion during quantum cooling is complex
with peaks at the diffusion fronts, enabling faster reaching of the answers similar to diffusion
in quantum walks (QWs) \cite{Childs2004,Childs2009,Venegas2012,Liu2023}. On the other hand, the diffusion during classical cooling is weaker, and the probability distribution decreases monotonically with Hamming distance
like the classical random walk (RW). These observations are depicted in Fig. \ref{fig:diffusion}(c).

Diffusion also occurs in the bath, leading to its state being driven away from its energy minimum. Better cooling performance is achieved when the bath is excited to higher energy states with greater probability and speed. Fig. \ref{fig:diffusion}(d) shows that the diffusion of the quantum bath is stronger with two peaks at the diffusion fronts, similar to the QW. On the other hand, the diffusion of the classical bath is weaker with monotonically decreasing fronts, as shown in Fig. \ref{fig:diffusion}(d). The classical bath diffuses very little without quantum coherence.

The diffusion in quantum cooling is similar to QW, as shown by the numerical results.
To further illustrate this similarity, we consider a one qubit  system
with $H_{\rm s}=\hbar\omega_{\rm s}\hat{\sigma}_z/2$ and $\omega_{\rm s}=\omega_{\rm b}=\omega$.
It is the Jaynes–Cummings model of non-rotating wave form \cite{Blais2021}. There are analytical results in the regime that one of the parameters $\lambda$, $\hbar\omega_{\rm s}$ or $\hbar\omega_{\rm b}$ is much smaller than the others \cite{Casanova2010}. The non-perturbative resonant regime $\lambda\sim\hbar\omega_{\rm s}=\hbar\omega_{\rm b}$ has rare analytical results. When the interaction is strong enough, an effective QW is observed
in the dynamics. For a small time interval $\Delta t$ \cite{Fernando2012}, its unitary evolution is (see Appendix \ref{app:QW} for derivation)
\begin{equation}
\label{eq:qw}
U = e^{-i\frac{H\Delta t}{\hbar}}=e^{-i\frac{H_0\Delta t}{\hbar}}U_pU_{\rm H}+O\left( \Delta t^3\right) \,,
\end{equation}
where the on-site energy is $H_0=H_{\rm s}+H_{\rm b}$. For a typical Hadamard walk in a one-dimensional space with a coin qubit,
its unitary operation of each iteration is $U_{\rm QW}=U_pU_{\rm H}$, where $U_{\rm H}$ is the Hadamard gate to the coin qubit and $U_p=e^{-\hat{\sigma}'_z\Delta x\partial/\partial x}$ is the conditional walking operator controlled by the coin qubit.
Therefore, in comparison,  our system has  an additional punishment term $e^{-i\frac{H_0\Delta t}{\hbar}}$ that may disturb the phase \cite{Childs2004}.
However, this  punishment term disappears when  the eigenstates of $H_0$ involved in the evolution are degenerate. This is exactly the
case that we have studied where  $H_{\rm s}$ and $H_{\rm b}$ are in resonance with matching energy gaps.
It is well known that
the average walking distance of RW is proportional to the square root of time $\Delta x\sim \sqrt{t}$
while that of QW is proportional to the time $\Delta x\sim t$ \cite{Venegas2012,Kempe2003}.
The constructive interference makes QW diffuse faster than RW.
It is also known that the entanglement between the coin qubit and the walking space is strong.

For the QW with multiple coin qubits, the situation is more complex (see Appendix \ref{app:QW}).
The analytical results always require non-local operation, such as the Grover's coin \cite{Carneiro2005,Shenvi2003}.
However, in our physical model, all the interactions are two-local. The numerical results in Fig. \ref{fig:diffusion} suggests that QW still makes contribution in this complex case.

\begin{figure}[tb]
\begin{center}
\includegraphics[clip = true, width =\columnwidth]{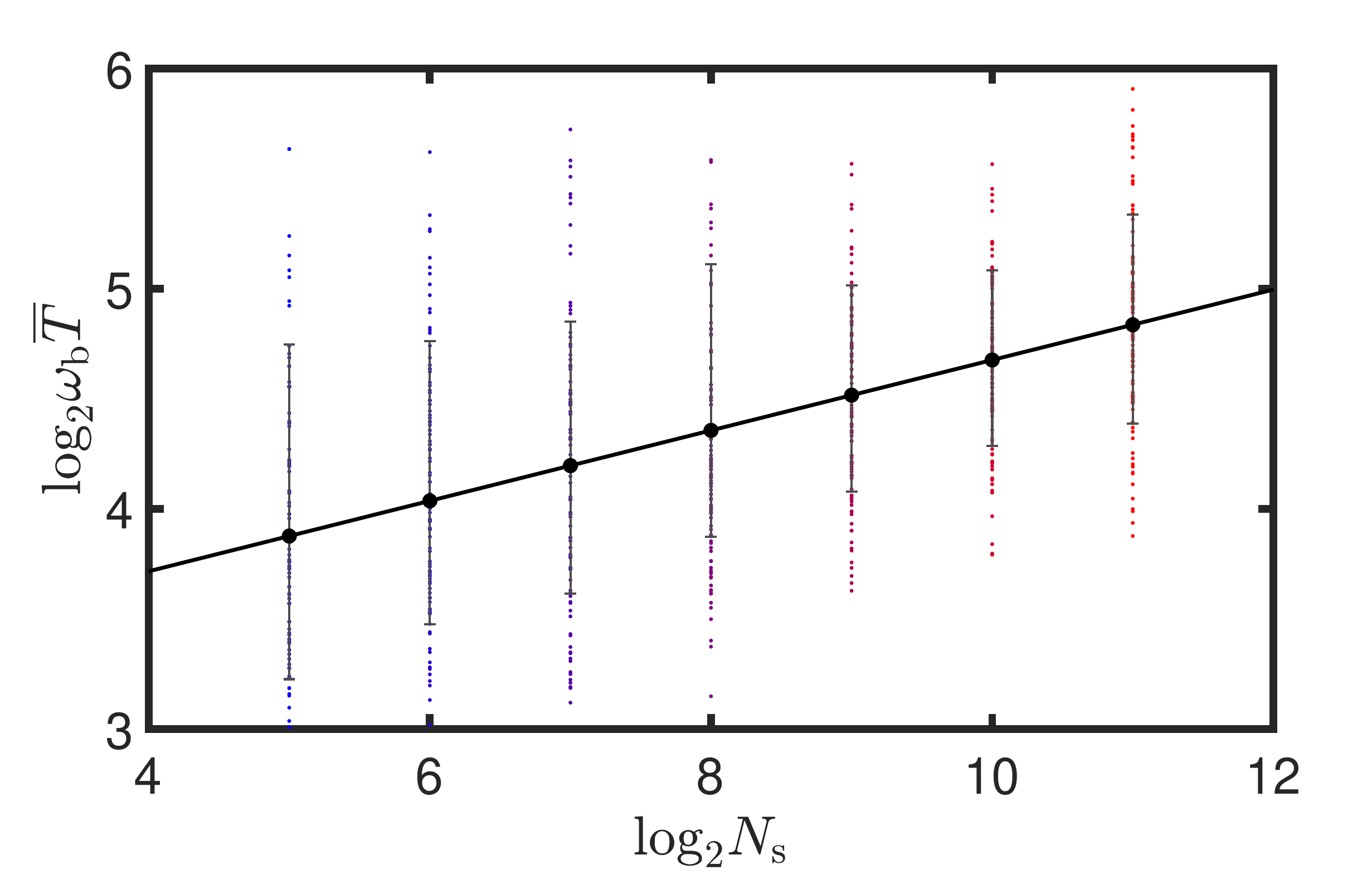}
\caption{\label{fig:TC} (color online) The average running time $\overline{T}$ against the size of Hilbert space $N_{\rm s}$. Each point represents a graph in samples. The solid line represents the regression line, which has a gradient of approximately 0.16. The error bars represent the standard deviation from the regression line, primarily resulting from the fluctuation of different graphs.}
\end{center}
\end{figure}

The quantum advancement of escaping local minima can be quantitatively approach. The numerical simulation of time complexity can be seen in Fig. \ref{fig:TC}, which the size of the Hilbert space is $N_{\rm s}=2^{n_{\rm s}}$. The time required for QIA to transition between local minima is approximately $O\left(N_{\rm s}^{0.16}\right)$ when solving specific local interaction problems (further details in Appendix \ref{app:TC}). There are $O\left(n_{\rm s}^{O(1)}\right)$ energy levels for the degenerate Hamiltonian which is proportional to the number of iterations and calls of the classical algorithm. So the total time complexity is about $O\left(N_{\rm s}^{0.16}n_{\rm s}^{O(1)}\right)$. Besides, there are connections to each qubit, so the total space complexity is $O\left(n_{\rm s}^2\right)$.

The concept of using a cavity for cooling has similarities to the quantum-circuit refrigerator (QCR) \cite{Tan2017,Silveri2017,Hsu2020}, which utilizes an open cavity to cool a system. However, the QCR exhibits strong dissipation in the cavity,
and some schemes utilize a dissipative qubit as the bath \cite{Meghana2020,Polla2021}.
Although an open bath can introduce steady energy loss, the lack of quantum coherence may hinder the cooling process.
Recent research suggests that the non-Markovian
baths could improve the performance of a quantum refrigerator \cite{Camati2020,Zhang2021}. In Appendix \ref{app:master}, we show that the dissipation in the quantum bath will slow down the cooling speed.

In conclusion, we proposed a hybrid quantum-classical algorithm of coherent cooling to address the issue of local minima.
It encodes the cost function in a superconductor circuit cooled by a quantum cavity,
which outperforms a classical cavity due to the coordinative tunneling effect.
Our results suggest that the faster diffusion of QW over classical RW is the underlying physics of this quantum speed-up. While we have focused on discrete variable problems, our scheme can be extended to continuous variable problems. This hybrid algorithm is expected to streamline the control and reduce noise in quantum computing experiments.

\bigskip

\begin{acknowledgments}
BW and JJF are supported by the National Key R\&D Program of China (Grants No.~2017YFA0303302, No.~2018YFA0305602), National Natural Science Foundation of China (Grant No. 11921005), and Shanghai Municipal Science and Technology Major Project (Grant No.2019SHZDZX01).
\end{acknowledgments}

\appendix
\section{Superconductor Circuit of Coherent Cooling\label{app:circuit}}

\begin{figure}[tb]
\begin{center}
\includegraphics[clip = true, width =\columnwidth]{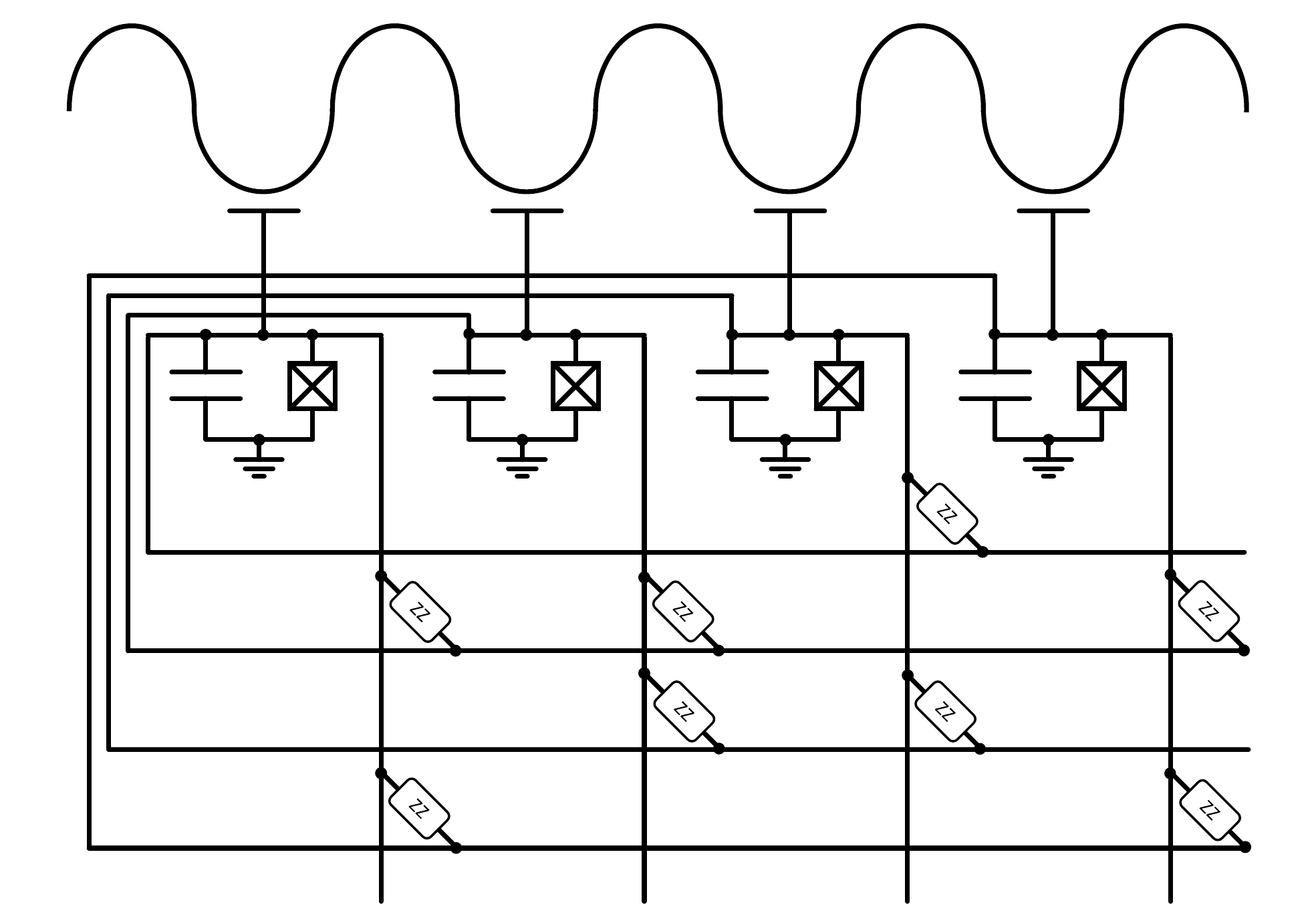}
\caption{\label{fig:circuit} The superconductor circuit of coherence cooling with the cavity. The long curve on the top is the cavity modeled by the $LC$ oscillator. The symbol with a cross inside the square is the Josephson junction, which constitute the qubit with the parallel capacitor. There is no connection of each intersection between wires without a dot.}
\end{center}
\end{figure}

Superconductor circuits can be utilized to physically implement the quantum icebox algorithm, and we can use the simplest charge qubits of Josephson junction as an example as shown in Fig. \ref{fig:circuit}. Other types of qubits require slight modifications. For phase qubits, an additional injected current is needed for each qubit. For flux qubits, the on-site capacitors are replaced by inductors, and the interconnected capacitors are replaced by mutual inductors \cite{Harris2009,Zhao2020,Dai2021}. The on-site energy $J_m^{(1)}$ can be tuned by the parallel capacitance of each qubit, while the interaction $J_{m,m'}^{(2)}$ depends on the connectivity and coupling strength of the circuit matrix. The $ZZ$ interaction is usually achieved by effective indirect interaction \cite{Brink2005,Collodo2020,Zhao2020}. If the coherence of long-distance interaction is poor, indirect interaction can be used to improve it \cite{Qiu2020,Zhong2019,Majer2007,Blais2007,Zhao2022,Kannan2020,Zhong2021}.
The crosstalk can be suppressed by a thick insulating layer made of high dielectric constant materials \cite{Zhao2022}.

The design shown in Fig. \ref{fig:circuit} is inspired by the matrix circuit, a design commonly used in classical circuits for devices like screens, keyboards, and memories. The information is encoded in the intersections of the rows and columns, and only two circuit layers are needed. IBM have already realized the multi-layer chip of quantum computer, while there are tens of layers used in modern classical CPUs.

\section{The Influence of Strong Interaction\label{app:SC}}

For quantum acceleration, strong interaction is required to overcome localization, which typically occurs in low-symmetry quantum systems and suppresses the propagation of wave functions \cite{Schreiber2011,Abanin2019}.

\begin{figure}[tb]
\begin{center}
\includegraphics[clip = true, width =\columnwidth]{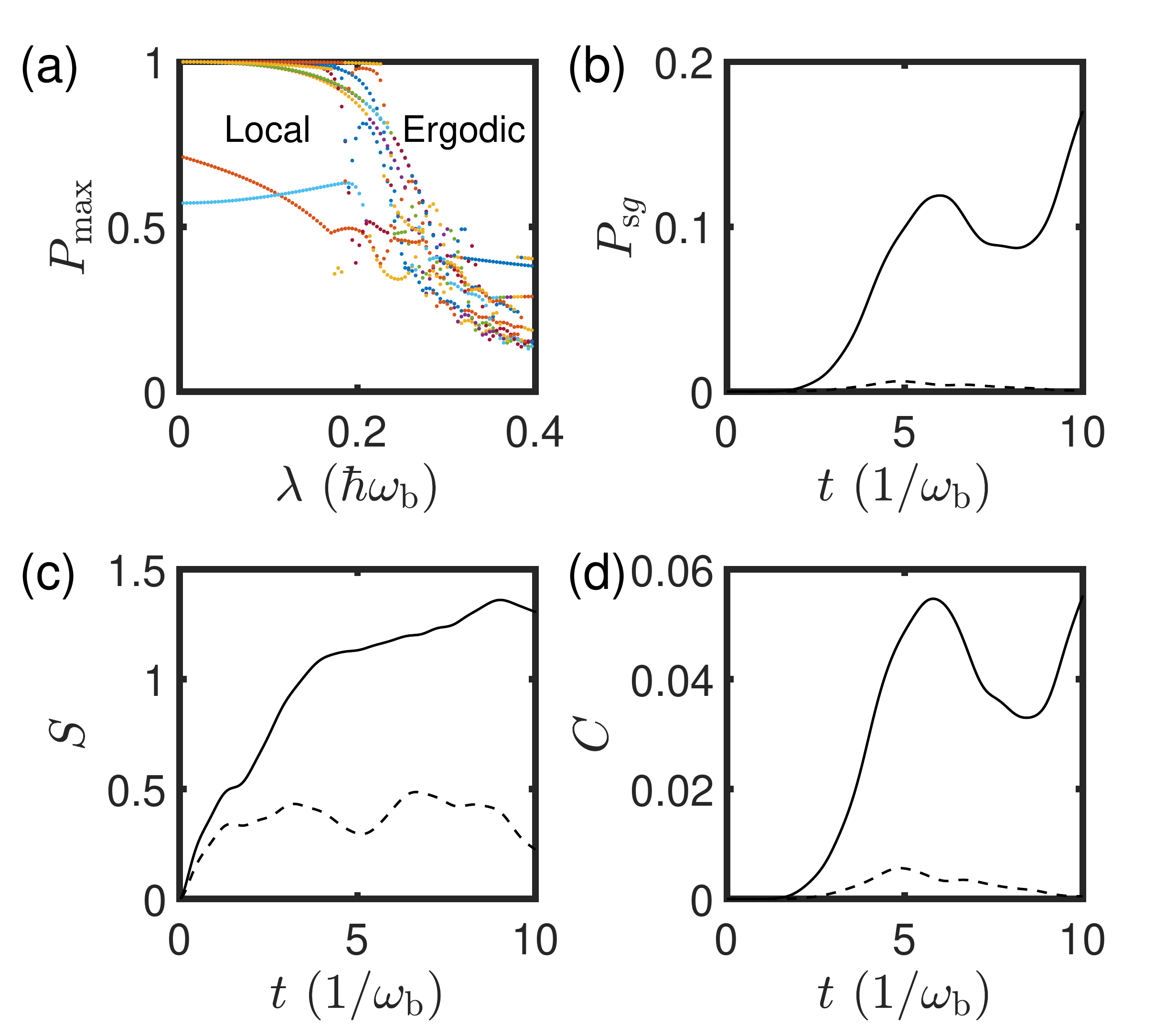}
\caption{\label{fig:S} (color online) (a) The localization strength of ten low energy eigenstates distinguished by different
colors. Time dependent evolution of (b) the ground state probability of the problem system, (c) the entanglement entropy, and (d) the correlation of join probability. The interaction strength is $\lambda=0.2\hbar\omega_{\rm b}$ (solid line) and $\lambda=0.15\hbar\omega_{\rm b}$ (dashed line).}
\end{center}
\end{figure}

We begin by examining the eigenstates of the total Hamiltonian $H=H_{\rm s}+H_{\rm b}+H_\lambda$.
In the absence of interaction ($H_\lambda$=0), the eigenstates are product states of the system and bath. The system is fully localized at the vertex of the hypercube in the $\sigma_z$ basis, with no nearby probability distribution. If localization decreases, probability distribution appears nearby, and the peak of probability $P_{\rm max}$ decreases as the probability is shared with other states. Thus, localization strength can be determined by $P_{\rm max}$ of eigenstates with low energy, shown in Fig. \ref{fig:S}(a). Weak interaction ($\lambda$) results in highly localized eigenstates \cite{Abanin2019}, as seen on the left side of Fig. \ref{fig:S}(a), where the wave function is unable to reach the whole Hilbert space, preventing solution discovery. On the other hand, Strong interaction causes delocalization \cite{Childs2004,Potter2015,Serbyn2015}, as shown on the right side of Fig. \ref{fig:S}(a), with sufficient probability distribution around the ground state.

Furthermore, the degree of localization can also impact the strength of revivals. Revivals typically occur in (pseudo-)integrable systems, where the state after evolution partially returns to its initial state. When the interaction is weak, only a few states are involved, resulting in strong revivals as illustrated by the dashed line in Fig. \ref{fig:S}(b). On the other hand, when the interaction is comparable to the on-site energy, more states are involved, leading to a rich energy spectrum of the initial state, and weaker revivals as shown by the solid line in Fig. \ref{fig:S}(b).


We use the entanglement entropy and correlation to further investigate the quantum coherence
between the quantum bath and the problem system. The entanglement entropy is given by
\begin{eqnarray}
S=-{\rm Tr}\left( \rho_{\rm s}{\rm ln}\rho_{\rm s} \right)~,
\end{eqnarray}
where $\rho_{\rm s}={\rm Tr}_{\rm b}\left( \rho \right)$ is the reduced density matrix of the problem system.
The entanglement entropy is shown in Fig. \ref{fig:S}(c).
The correlation can be described by the joint probability defined as \cite{Wen2019}
\begin{eqnarray}
C&=&\left\langle \hat{P}_{{\rm s}g}\left(\hat{I}-\hat{P}_{{\rm b}g}\right)\right\rangle-\left\langle \hat{P}_{{\rm s}g}\right\rangle\left\langle  \hat{I}-\hat{P}_{{\rm b}g}\right\rangle \nonumber\\
&=&-\left\langle \hat{P}_{{\rm s}g}\hat{P}_{{\rm b}g}\right\rangle+\left\langle \hat{P}_{{\rm s}g}\right\rangle\left\langle  \hat{P}_{{\rm b}g}\right\rangle \,.
\end{eqnarray}
Fig. \ref{fig:S}(d) shows the correlation.

When the interaction between the problem system and the bath is weak ($\langle H_\lambda\rangle \ll \langle H_0\rangle$), the bath's influence can be considered a perturbation. Thus, the composed system can be approximated as a product state of the problem system and the bath, and the rotating wave approximation (RWA) holds to a large extent. Both the entanglement entropy and correlation are small, as shown by the dashed lines in Figure \ref{fig:S}. In this scenario, the localization is strong, but the cooling efficiency is low \cite{Abanin2019}.

As the interaction strength becomes comparable with the on-site energy, $\langle H_\lambda\rangle\sim \langle H_0\rangle$,
the composed system undergoes a phase transition \cite{Luitz2015} and the wave function becomes delocalized and ergodic \cite{Potter2015,Serbyn2015}. The product state approximation of the problem system
and the bath is no longer valid. The entanglement and correlation between them increase significantly
as depicted by the solid lines in Fig. \ref{fig:S}, while the entanglement between the system and the classical bath is absent.

At the extremely strong coupling regime, the interaction $H_\lambda$ dominates the dynamics. Despite the strong entanglement and correlation, the information from the system with $H\approx H_\lambda$ is limited. Consequently, the cooling effect will vanish, leading to a low success rate.

\section{Cooling the Continuous System\label{app:Continous}}

\begin{figure}[tb]
\includegraphics[clip = true, width =\columnwidth]{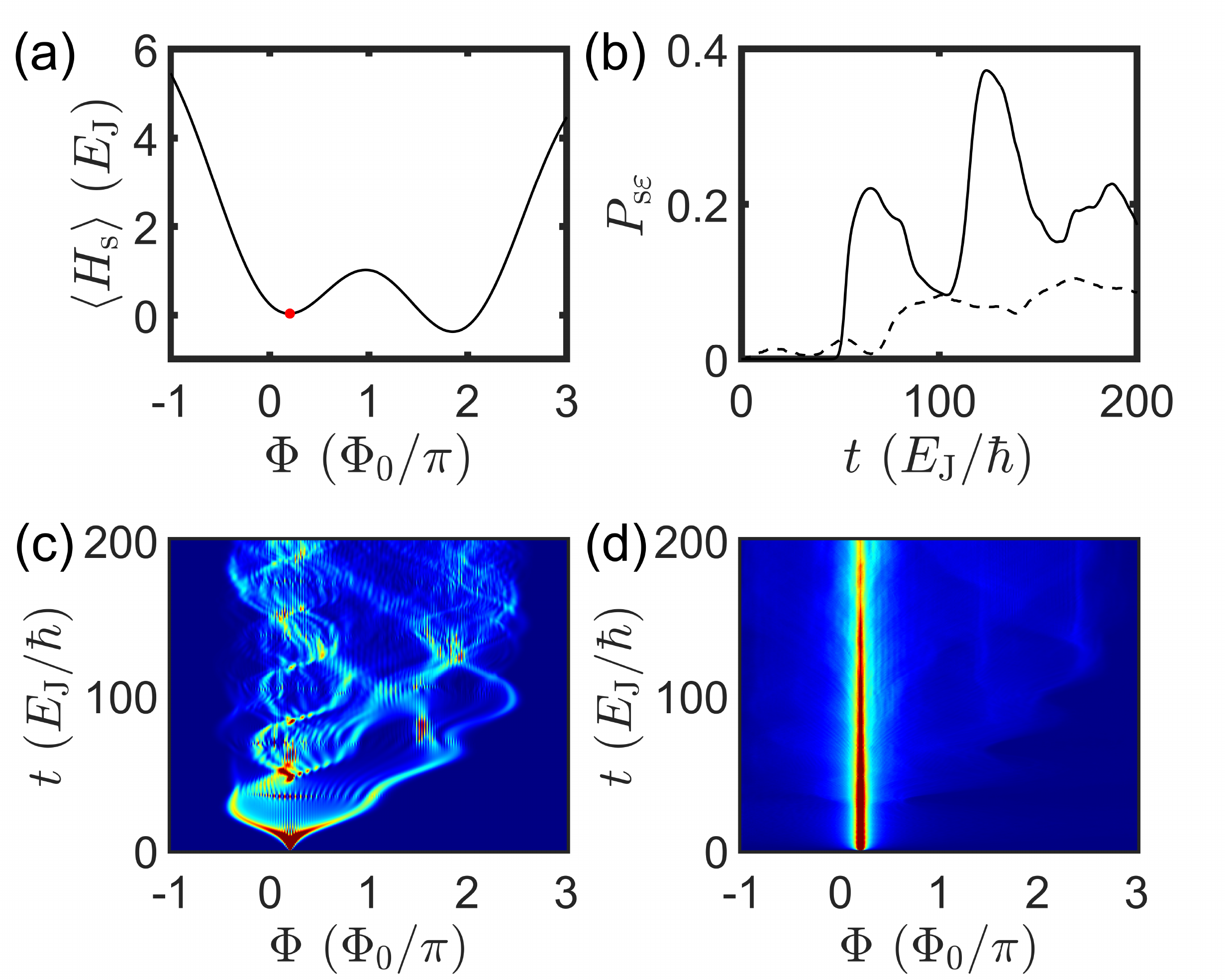}
\caption{\label{fig:pq} (color online) (a) The system energy as a function of $\Phi$; the red point indicates the initial state. (b) The time evolution of the total probability of the states which have lower on-site energy than the initial state. The solid line is for cooling using the quantum bath while the dashed line is cooling using the classical bath. (c) The time evolution of the probability distribution of the system with (c) the quantum bath and (d) the classical bath (red for high values and blue for low).}
\end{figure}

For problems with continuous dynamical variables,  they can also be encoded in Hamiltonians,
for example, by mapping the variables to the generalized positions $\Phi_{\rm s}$ of Josephson junctions.
The cost function for such a problem can be encoded in a Hamiltonian like this:
\begin{eqnarray}
H_{\rm s} &=& \frac{\left(\hat{\Phi}_{\rm s}-\Phi_{\rm ex}\right)^2}{2L_{\rm s}}-E_{\rm J}\cos\frac{2\pi\hat{\Phi}_{\rm s}}{\Phi_0} ~,
\end{eqnarray}
where $\Phi_{\rm ex}$ is the external magnetic flux, $\Phi_0$ is the quantum magnetic flux,
$L_{\rm s}$ is the inductance, and $E_{\rm J}$ is the Josephson energy. These parameters
can all be tuned \cite{Harris2010}. The goal is to find  the global minimum on $\Phi_{\rm s}$,
with the shape of the cost function shown in Fig. \ref{fig:pq}(a).
The initial state is set at the higher local minimum.

The system-bath interaction can arise via the capacitor connection.
This interaction can be described by the Hamiltonian
\begin{eqnarray}
H_{\lambda} &=& -\frac{\hat{Q}_{\rm s}\hat{Q}_{\rm b}}{C_\lambda} ,
\end{eqnarray}
where $C_\lambda$ is the effective capacitance of the connection. The influence of the on-site capacitance
has already been included in parameters of $H_{\rm s}$ and $H_{\rm b}$.

Cooling process with the quantum bath satisfies the Schr\"odinger equation.
Cooling process using the classical bath satisfies the following equations
\begin{eqnarray}
    i\hbar\frac{\partial\left| \psi_{\rm s} \right\rangle}{\partial t}&=&\left(H_{\rm s}-\frac{Q_{\rm b}}{C_\lambda}\hat{Q}_{\rm s} \right) \left| \psi_{\rm s} \right\rangle ,  \\
    \frac{\partial Q_{\rm b}}{\partial t}&=&-\frac{\Phi_{\rm b}}{L_{\rm b}} ,\\
     \frac{\partial \Phi_{\rm b}}{\partial t}&=&\frac{Q_{\rm b}}{C_{\rm b}} -\frac{\left\langle\hat{Q}_{\rm s}\right\rangle}{C_\lambda} .
\end{eqnarray}

Comparing the time evolutions in Fig. \ref{fig:pq}(b), it is observed that the quantum bath yields
better cooling efficiency than the classical bath in the continuous variable scenario.
The quantum cooling displays characteristic quantum oscillations, while classical cooling is smoother.
Fig. \ref{fig:pq}(c) illustrates the coherent diffusion of quantum cooling through coordinative
tunneling in the position space, displaying strong coherent enhancement at the wave function's fronts.
Fig. \ref{fig:pq}(d) shows the smoother incoherent diffusion of classical cooling,
with the strongest component trapped in the higher local minimum.

\section{Quantum Walk Effect in the Coherent Cooling\label{app:QW}}

Governed by the Schr\"{o}dinger equation, the time evolution of the composed Hamiltonian
can be divided into small steps
\begin{eqnarray}
e^{-i\frac{H t}{\hbar}}&=& \prod_{n=1}^{t/\Delta t} e^{-i\frac{H\Delta t}{\hbar}}\,.
\end{eqnarray}
The unitary operator for each step, $U=e^{-i\frac{H\Delta t}{\hbar}}$ for the Hamiltonian in Eq. (9) in the main text, can be approximated as \cite{Fernando2012}
\begin{eqnarray}
U&=&e^{-i\frac{\omega}{2}\hat{\sigma}_{z}\Delta t-i\omega\left(\hat{q}_{\rm b}^{2}+\hat{\phi}_{\rm b}^{2}\right)\Delta t+i2\frac{\lambda}{\hbar}\hat{\sigma}_{y}\hat{q}_{\rm b}\Delta t}\nonumber\\
&\thickapprox&e^{-i\frac{\omega}{2}\hat{\sigma}_{z}\Delta t-i\omega\left(\hat{q}_{\rm b}^{2}+\hat{\phi}_{\rm b}^{2}\right)\Delta t}e^{i2\frac{\lambda}{\hbar}\hat{\sigma}_{y}\hat{q}_{\rm b}\Delta t} \nonumber\\
&&e^{-\frac{1}{2}\left[-i\frac{\omega}{2}\hat{\sigma}_{z}\Delta t,i2\frac{\lambda}{\hbar}\hat{\sigma}_{y}\hat{q}_{\rm b}\Delta t\right]}
e^{-\frac{1}{2}\left[-i\omega\left(\hat{q}_{\rm b}^{2}+\hat{\phi}_{\rm b}^{2}\right)\Delta t,i2\frac{\lambda}{\hbar}\hat{\sigma}_{y}\hat{q}_{\rm b}\Delta t\right]} \nonumber\\
&=&e^{-i\frac{\omega}{2}\hat{\sigma}_{z}\Delta t-i\omega\left(\hat{q}_{\rm b}^{2}+\hat{\phi}_{\rm b}^{2}\right)\Delta t}
e^{i2\frac{\lambda}{\hbar}\hat{\sigma}_{y}qt}e^{i\frac{\lambda}{\hbar}\omega\hat{\sigma}_{x}\hat{q}_{\rm b}\Delta t^{2}}
e^{-i\omega\frac{\lambda}{\hbar}\hat{\sigma}_{y}\hat{\phi}_{\rm b}\Delta t^{2}}\nonumber\\
&\thickapprox&e^{-i\frac{\omega}{2}\hat{\sigma}_{z}\Delta t-i\omega\left(\hat{q}_{\rm b}^{2}+\hat{\phi}_{\rm b}^{2}\right)\Delta t}e^{i\frac{\lambda}{\hbar}\left(2\Delta t\hat{\sigma}_{y}+\omega \Delta t^{2}\hat{\sigma}_{x}\right)\hat{q}_{\rm b}}
e^{-i\omega\frac{\lambda}{\hbar}\hat{\sigma}_{y}\hat{\phi}_{\rm b}\Delta t^{2}} ~, \nonumber\\
\end{eqnarray}
where we have omitted all the $O\left( \Delta t^3\right)$ terms. The dimensionless operator is $\hat{q}_{\rm b}=-i\partial/\partial \phi_{\rm b}$. Notice that the angle of rotation axis between $U_p$ and $U_{\rm H}$ of the coin qubit is $\pi/4$ in the Hadamard walk. In the time interval $\Delta t=2/\omega$, this condition will be satisfied with
\begin{eqnarray}
U &=&e^{-i\frac{H_0\Delta t}{\hbar}}
e^{-\frac{2\sqrt{2}\lambda}{\hbar\omega}\frac{\hat{\sigma}_{y}+\hat{\sigma}_{x}}{\sqrt{2}}\frac{\partial}{\partial \phi_{\rm b}}}
e^{-i\frac{4\lambda}{\hbar\omega}\hat{\sigma}_{y}\hat{\phi}_{\rm b}} ~.
\end{eqnarray}
Then the rotation axis between the last two terms is also $\pi/4$. To simplify the comparison with the Hadamard walk, we do the transformation that
\begin{eqnarray}
\hat{\sigma}_{z}&=&\hat{\sigma}_{x}' ~,\nonumber\\
\hat{\sigma}_{x}&=&\frac{\hat{\sigma}_{z}'-\hat{\sigma}_{x}'}{\sqrt{2}} ~,\nonumber\\
\hat{\sigma}_{y}&=&\frac{\hat{\sigma}_{z}'+\hat{\sigma}_{x}'}{\sqrt{2}} ~.
\end{eqnarray}
Then the operation  becomes
\begin{eqnarray}
U =e^{-i\frac{H_0\Delta t}{\hbar}} e^{-\hat{\sigma}'_{z}\frac{2\sqrt{2}\lambda}{\hbar\omega}\frac{\partial}{\partial \phi_{\rm b}}}e^{i\frac{4\lambda}{\hbar\omega}\frac{\hat{\sigma}'_{z}+\hat{\sigma}'_{x}}{\sqrt{2}}\hat{\phi}_{\rm b}} \,.
\end{eqnarray}
When $4\lambda\phi_{\rm b}/\hbar\omega=\left(2n+1\right)\pi$, namely,
$\Delta \phi_{\rm b}=\pi_{\rm b}\hbar\omega/2\lambda$, the last term is  the Hadamard gate. Moreover, the displacement of conditional translation is $\Delta \phi_{\rm b}=2\sqrt{2}\lambda/\hbar\omega$, so the self-consistent requires $\lambda=\sqrt{\pi}/2^\frac{5}{4}\hbar\omega$. In summary, the unitary operation is
\begin{eqnarray}
U &=&e^{-i\frac{H_0\Delta t}{\hbar}}U_pU_{\rm H}\,.
\end{eqnarray}
Except the  term $e^{-i\frac{H_0\Delta t}{\hbar}}$, it is just the  standard Hadamard walk.

In the case of spin glass with multi-qubits, the operation is
\begin{eqnarray}
U &=&e^{-i\frac{H_0\Delta t}{\hbar}}U_pU_{\rm H}^{\otimes n_{\rm s}} ~,
\end{eqnarray}
where the Hadamard gates require $\omega_{\rm b}\lambda\Delta\phi_{\rm b}\Delta t^2/\hbar=2\pi$. The conditional walker is more complex that
\begin{eqnarray}
U_p = e^{i\frac{\lambda}{\hbar}\left(\sum_{m=1}^{n_{\rm s}}\Delta t\hat{\sigma}_m^y+\sum_{m=1}^{n_{\rm s}}J_m^{(1)}\Delta t^2\hat{\sigma}_m^x
+\sum_{\langle m,m'\rangle}J_{m,m'}^{(2)}\Delta t^2( \hat{\sigma}_m^x\hat{\sigma}_{m'}^z+\hat{\sigma}_m^z\hat{\sigma}_{m'}^x )  \right)\frac{\partial}{\partial \phi_{\rm b}}}  ~.\nonumber\\
\end{eqnarray}

\section{Time Complexity\label{app:TC}}

In this section, we will attempt to evaluate the time complexity of our algorithm. Our algorithm is general and applicable to many optimization problems. However, in order to gain informative results with limited computational resources, we need to narrow our focus to a specific set of problems.
These optimization problems are constructed in the following manner. All qubits are first connected in a one-dimensional chain to make them inseparable. We then connect each qubit to another randomly selected qubit. The number of links for each qubit ranges from 2 to $n_{\rm s}+2$, with an average of $2-(0.5/n_{\rm s})$ links per qubit.
Ferromagnetic interaction exists only between two connected qubits with equal strength. As a result, there are two global minima $|0\rangle^{\otimes n_{\rm s}}$ and $|1\rangle^{\otimes n_{\rm s}}$. Finally, we add on-site energy to certain qubits. The two ground states will split:
one becomes a local minimum with high-energy
and the other remains a global minimum.

For a system with $n_{\rm s}$ qubits, it can correspond to a graph. As there are
$O\left(2^{n_{\rm s}^2}\right)$ different graphs, and the configuration space is extremely large.
It is  difficult to traverse all of the graphs. As a compromise, we sample one hundred graphs following the mentioned rule for each value of $n_{\rm s}$ and consider them as typical.

\begin{figure}[tb]
\begin{center}
\includegraphics[clip = true, width =\columnwidth]{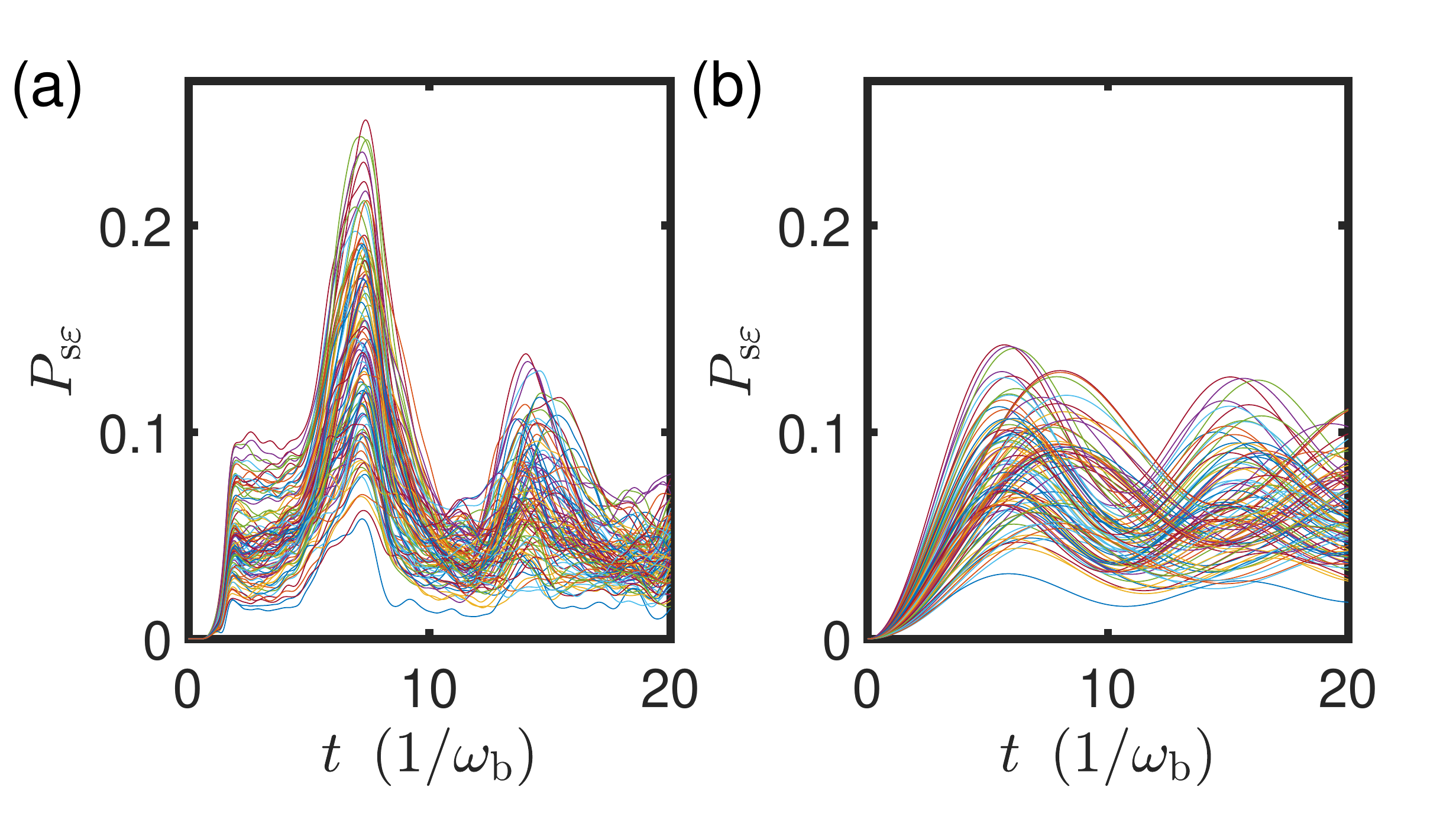}
\caption{\label{fig:fit} (color online) (a) The time-dependent total probability of states that have lower on-site energy than the initial state. Different colors represent different graphs. The size of the problem system is $n_{\rm s}=11$. (b) The fitting curves of (a) with just two adjustable parameters. }
\end{center}
\end{figure}

If the problem system reaches the global minimum, we consider the cooling process successful, which can be described by the characteristic time $\tau$ and the success rate $P_{\rm r}$.
The characteristic time $\tau$ refers to the speed at which the dynamics reach equilibrium. The success rate $\tau$ is the total probability at the global minimum when the dynamics reach equilibrium.
To extract the characteristic time and success rate, we
fit the numerical results with the following function
\begin{eqnarray}
P_{{\rm s}\varepsilon} &\approx& P_{\rm r}f\left( \frac{t}{\tau}\right) ~,
\end{eqnarray}
where the function has the properties that ${\rm lim}_{x \rightarrow+\infty}f(x)=1$ and ${\rm lim}_{x \rightarrow0}f(x)/x^2=constant$. Using the fitting function can suppress the poisoning of the results caused by accidental narrow peaks and outliers. Here, we have chosen the fitting function to be $f(x)=1-J_0(x)$, where $J_0(x)$ is the zero-order Bessel function of the first kind. The fitting performance is shown in Fig. \ref{fig:fit}. Since there are only two free parameters, the fitting may not be perfect. Other fitting functions are tried, yielding parameters of similar magnitude.

Now, we will begin analyzing the time complexity. The average running time can be defined as
\begin{eqnarray}
\overline{T} &=& \frac{\tau}{P_{\rm r}-\frac{1}{N_{\rm s}}} ~.
\end{eqnarray}
where $\tau$ is the characteristic time of the dynamics and $P_{\rm r}$ is the success rate. $\overline{T}$ is approximately proportional to the runtime required to achieve a given success rate. If  $P_{\rm r}$ is smaller, more attempts and measurements will be required. If the interaction is exceedingly strong, the system will rapidly become ergodic. However, there is no calculation associated with this type of probability increment. So, we subtract the background probability of such a wild guess, which is $1/N_{\rm s}=2^{-n_{\rm s}}$. The average running time against the system's size is shown in Fig. \ref{fig:TC}. The best and worst cases are not shown because they are difficult to sample with a large value of $n_{\rm s}$. For the average case in samples, the regression yields a gradient of approximately 0.16. The time complexity for each successful cooling is $O\left( N_{\rm s}^{0.16} \right)$.

The encoding Hamiltonian is highly degenerate with $O\left(n_{\rm s}^{O(1)}\right)$ energy levels. The  maximum number of iterations required to reach the lowest energy levels is also bounded by $O\left(n_{\rm s}^{O(1)}\right)$. Specifically, it is $O\left(n_{\rm s}^3\right)$ for solving the maximal independent set problem with the two-local Hamiltonian.

In summary, the total average time complexity is approximately $O\left(N_{\rm s}^{0.16}{\rm log}_2^3N_{\rm s}\right)=O\left(1.12^{n_{\rm s}}n_{\rm s}^3\right)$. The time complexity of the classical algorithm is approximately $O\left(1.19^{n_{\rm s}}n_{\rm s}^{O(1)}\right)$ \cite{Mingyu2017}.

\section{Decoherence with Markov Master Equation\label{app:master}}
\begin{figure}[tb]
\begin{center}
\includegraphics[clip = true, width =\columnwidth]{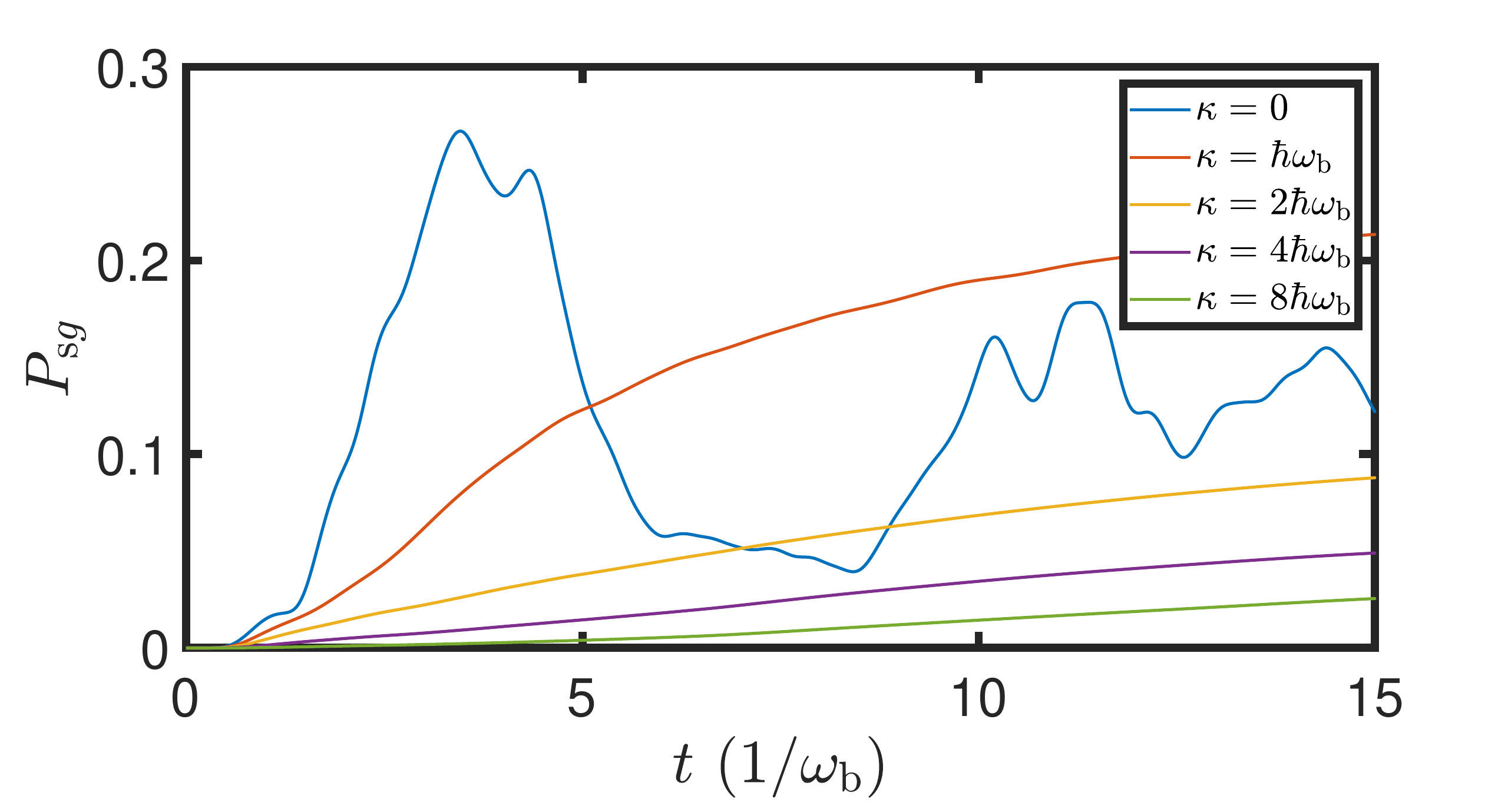}
\caption{\label{fig:decoherence} (color online) The ground state probability of the problem system with different decoherence strength $\kappa$ of the bath.
Other parameters are $n_{\rm s}=5$ and $\lambda=0.6\hbar\omega_{\rm b}$.}
\end{center}
\end{figure}

In the main text, we have discussed cooling with a coherent quantum bath using the Schr\"{o}dinger equation. However, the experiment always faces decoherence, such as the thermal bath, which has been shown to easily reach local minima \cite{chen2023}.
Now we investigate the ability to reach the global minimum and treat the cooling with the Markov master equation \cite{Orszag2001,li2023}
\begin{eqnarray}
\frac{{\rm d} \rho}{{\rm d} t}=-\frac{i}{\hbar}[H,\rho]- \frac{\kappa}{\hbar} \left( b^\dagger b\rho+ \rho b^\dagger b-2b\rho b^\dagger\right) ~,
\label{eq:master}
\end{eqnarray}
where $\kappa$  represents the decoherence strength of the bath.
 In Fig. \ref{fig:decoherence}, we compare the cooling processes with different values of $\kappa$.
 The case with $\kappa=0$ corresponds to the case discussed in the main text.

The numerical results indicate that the cooling process without decoherence is faster in the early stage. Although decoherence can extract energy from the bath irreversibly, it can disrupt the quantum acceleration by suppressing the off-diagonal terms of the density matrix $\rho$. This transition between quantum states requires these off-diagonal terms, and the present of decoherence can hinder quantum tunneling.

At the late cooling stage, the cooling process reaches saturation. Even a small amount of decoherence can decrease the revival from quantum oscillation \cite{Kendon2003,Schreiber2011}. Strong decoherence, on the other hand, can entirely suppress the quantum transition.

We consider a simple case to demonstrate the effect of decoherence on quantum acceleration,
where both the problem system and the bath are two-level systems \cite{Camati2020,Zhang2021}.
The states $|\sigma_z, n_{\rm b}\rangle$ are re-defined as $|2\rangle\equiv|1,0\rangle$, $|1\rangle\equiv|-1,1\rangle$, and $|0\rangle\equiv|-1,0\rangle$.
Eq. (\ref{eq:master}) in RWA can then be written as
\begin{eqnarray}
\frac{{\rm d} \rho_{22}}{{\rm d} t} &=& \frac{\lambda}{\hbar}(i\rho_{12}-i\rho_{21}) ~, \\
\frac{{\rm d} \rho_{12}}{{\rm d} t} &=& \frac{\lambda}{\hbar}(i\rho_{22}-i\rho_{11})-\frac{\kappa}{\hbar}\rho_{12} ~, \\
\frac{{\rm d} \rho_{11}}{{\rm d} t} &=& \frac{\lambda}{\hbar}(-i\rho_{12}+i\rho_{21})-2\frac{\kappa}{\hbar}\rho_{11} ~, \\
\frac{{\rm d} \rho_{00}}{{\rm d} t} &=& 2\frac{\kappa}{\hbar}\rho_{11} ~,
\end{eqnarray}
where $\rho_{12}^\dagger=\rho_{21}$. The initial condition is $\rho_{22}=1$ and other elements of the density matrix are zero. The solution is
\begin{eqnarray}
 \rho_{22} &=& e^{-\frac{\kappa t}{\hbar}} \Big( \frac{1+\cos2\Omega t}{2}\frac{\lambda^2}{\hbar^2\Omega^2}
\nonumber\\
&& -\frac{\cos2\Omega t}{4}\frac{\kappa^2}{\hbar^2\Omega^2}+\frac{\sin2\Omega t}{4}\frac{\kappa}{\hbar\Omega}\Big)\,,
\end{eqnarray}
where $\hbar\Omega=\sqrt{\lambda^2-\kappa^2/4}$ and $P_{{\rm s}g}=1-\rho_{22}=\rho_{11}+\rho_{00}$.
It is a  damped oscillator with the decoherence $\kappa$ slowing down the oscillation angular frequency
from $\lambda/\hbar$ to $\Omega$. As a result, the quantum transition between $|2\rangle$ and $|1\rangle$
is also slowed down.

When the decoherence is small $\kappa\ll\lambda$, the system is in the underdamped regime with oscillations. The probability of being in the ground state is approximately given by
\begin{eqnarray}
P_{{\rm s}g}&\approx&1-e^{-\frac{\kappa t}{\hbar}} \left( \frac{1+\cos\frac{2\lambda t}{\hbar}}{2}+\frac{\kappa}{2\lambda}\sin\frac{2\lambda t}{\hbar} \right) ~.
\label{eq:smalldeco}
\end{eqnarray}
In the early stage $t\ll\hbar/\lambda$, Eq. (\ref{eq:smalldeco}) can be simplified to
\begin{eqnarray}
P_{{\rm s}g}&\approx&\sin^2\frac{\lambda t}{\hbar}-\frac{\lambda^2\kappa}{2\hbar^3}t^3 ~.
\end{eqnarray}
The last term is due to decoherence, and its negativity implies that it leads to deceleration.

When $\kappa\gg\lambda$, the system is in the overdamped regime without oscillations,
and the ground state probability is given by
\begin{eqnarray}
P_{{\rm s}g}\approx1-e^{-\frac{2\lambda^2}{\hbar\kappa}t} \,,
\label{eq:Pexp}
\end{eqnarray}
which indicates that stronger decoherence leads to slower cooling. In this regime,
the bath is almost frozen in its ground state, and the dynamics are essentially static.

For multi-qubits, as depicted in Fig. \ref{fig:decoherence}, there is also exponential behavior observed in the presence of strong dissipation, similar to Eq. (\ref{eq:Pexp}). Therefore, to ensure reasonable quantum acceleration, the decoherence strength should not greatly exceed the tunneling strength during quantum computing.

\normalem

\end{document}